\documentclass[compactaffiliation]{Interspeech}

\usepackage{multirow}
\usepackage{graphicx}
\usepackage{subcaption}
\usepackage{tikz}
\usepackage{cite}
\usepackage{hyperref}
\usepackage{diagbox}

\tikzstyle{mybox} = [draw=black, very thick,
    rectangle, rounded corners, inner sep=10pt, inner ysep=13pt]
\tikzstyle{fancytitle} =[fill=black, text=white]

\interspeechcameraready

\title{VocalAgent: Large Language Models for Vocal Health Diagnostics \\ with Safety-Aware Evaluation}

\author[affiliation={1}]{Yubin}{Kim}
\author[affiliation={5}]{Taehan}{Kim}
\author[affiliation={1}]{Wonjune}{Kang}
\author[affiliation={1}]{Eugene}{Park}
\author[affiliation={2}]{Joonsik}{Yoon}
\author[affiliation={3}]{Dongjae}{Lee}
\author[affiliation={4}]{\\Xin}{Liu}
\author[affiliation={4}]{Daniel}{McDuff}
\author[affiliation={2}]{Hyeonhoon}{Lee}
\author[affiliation={1}]{Cynthia}{Breazeal}
\author[affiliation={1}]{Hae Won}{Park}

\affiliation{}{Massachusetts Institute of Technology, }{USA}
\affiliation{}{Seoul National University Hospital, }{South Korea}
\affiliation{}{Doctor Diary, }{South Korea}
\affiliation{}{Google Research, }{USA}
\affiliation{}{Independent Researcher}{}
\email{ybkim95@mit.edu}


\begin{document}

\maketitle

\begin{abstract}
Vocal health plays a crucial role in peoples' lives, significantly impacting their communicative abilities and interactions. However, despite the global prevalence of voice disorders, many lack access to convenient diagnosis and treatment. This paper introduces VocalAgent, an audio large language model (LLM) to address these challenges through vocal health diagnosis. We leverage Qwen-Audio-Chat fine-tuned on three datasets collected in-situ from hospital patients, and present a multifaceted evaluation framework encompassing a safety assessment to mitigate diagnostic biases, cross-lingual performance analysis, and modality ablation studies. VocalAgent demonstrates superior accuracy on voice disorder classification compared to state-of-the-art baselines. Its LLM-based  method offers a scalable solution for broader adoption of health diagnostics, while underscoring the importance of ethical and technical validation.
\end{abstract}

\section{Introduction}


Voice disorders are a widespread global health concern, impacting millions and significantly affecting quality of life, professional opportunities, and relationships. Global lifetime prevalence is estimated to be around 29.1\% \cite{claros2019psychogenic}. In the United States, the problem remains significant, with approximately 1 in 5 adults (20.6\%) reporting they have experienced a voice disorder at some point in their lives\cite{huston2024prevalence}. Despite this considerable prevalence, access to timely diagnosis, effective treatment, and long-term management for these disorders remains a challenge, due to disparities influenced by factors such as socioeconomic status and race, particularly in underserved regions \cite{rameau2023addressing}.

Vocal health assessments typically require specialized equipment and expertise, making individualized treatment plans difficult to scale~\cite{asha_voice_disorders}. These challenges often result in delayed diagnoses and inconsistent care. While machine learning models for diagnosing vocal health show potential~\cite{rehman2024voice}, they primarily focus on binary classification (presence/absence of disorders) and largely lack explainability or personalization. This highlights the need for frameworks that enable practical, clinician-patient-AI collaboration in this field~\cite{bansal2024challenges, tu2024towards}.

Meanwhile, advances in deep learning have shown promise for various speech and voice analysis tasks. Self-supervised representation learning models such as wav2vec~2.0~\cite{baevski2020wav2vec}, HuBERT~\cite{hsu2021hubert}, and WavLM~\cite{chen2022wavlm} excel in tasks such as speech recognition~\cite{zhao2022improving}, speaker identification~\cite{lin2024sa}, and emotion recognition~\cite{wang2021fine}, and have also demonstrated some success in vocal health diagnostics~\cite{zhu2024wavrx, cai2024voice, cui2023transferring}.
More recently, there has been a wide range of research into augmenting large language models (LLMs)~\cite{achiam2023gpt, team2023gemini} with speech and audio understanding abilities~\cite{chu2023qwen, hu2024wavllm, gong2024listen}, thereby obtaining comprehensive audio understanding in diverse domains, including medical diagnostics. These systems offer the potential for providing explainable insights and personalized treatment strategies, bridging critical gaps in healthcare accessibility and care practice~\cite{alsaad2024multimodal, kim2024health}.

In this paper, we introduce \textbf{VocalAgent}, a speech LLM framework designed to enhance vocal health diagnosis.  Traditional machine learning approaches for voice disorder classification often focus on binary decisions and lack the nuanced understanding and explainability required for effective clinical use.  Moreover, they frequently fall short in providing interpretable insights and personalized assessments, hindering practical human-AI collaboration.  To address these limitations, VocalAgent leverages the power of speech LLMs to enhance vocal health diagnosis and management by offering more granular analysis and interpretable outputs.
Specifically, we fine-tune Qwen-Audio-Chat~\cite{chu2023qwen} for vocal health applications using in-situ patient data. Our key contributions are as follows:

\begin{enumerate}
    \item We present \textbf{VocalAgent}, a speech LLM-based framework for voice disorder classification that outperforms previous state-of-the-art models, as demonstrated on multiple datasets.
    \item We provide a set of comprehensive \textbf{safety-aware evaluations} that go beyond standard performance metrics. They collectively demonstrate the robustness and ethical considerations of VocalAgent for real-world clinical applications.
\end{enumerate}

\begin{figure*}[h]
    \centering
    \includegraphics[width=0.9\linewidth]{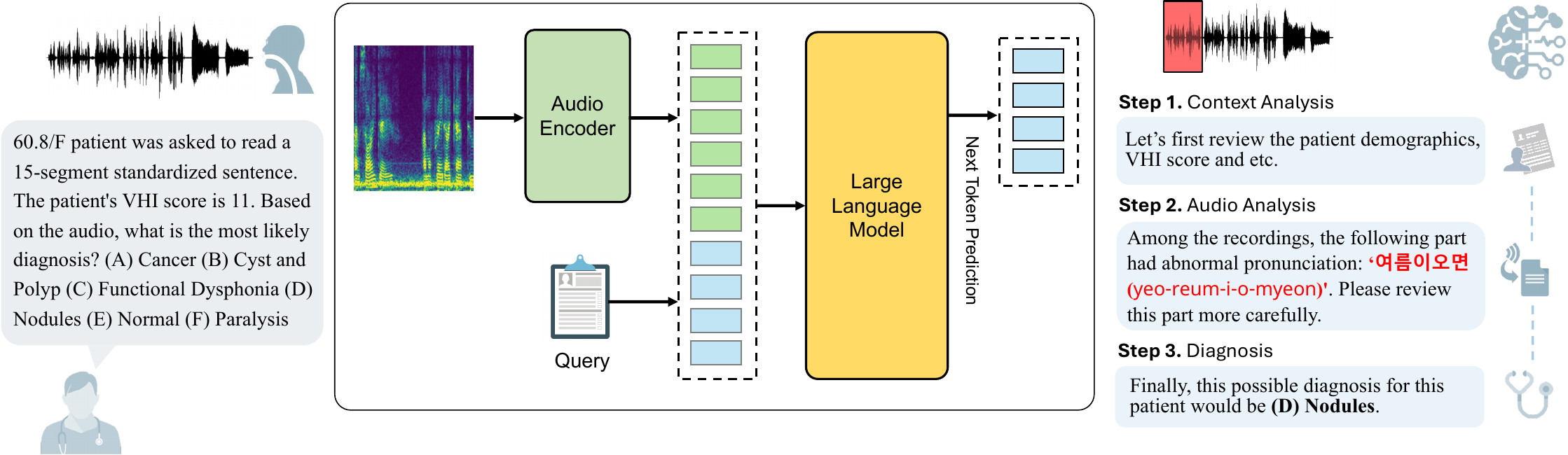} 
    \caption{\textbf{Schematic diagram of VocalAgent framework.} Based on patient audio, Voice Handicap Index (VHI) score, Grade, Roughness, Breathiness, Asthenia, Strain (GRBAS) scale and demographics, we conduct instruction fine-tuning on a base LLM.}
    \label{fig:main} 
    \vspace{-4pt}
\end{figure*}


\noindent \textbf{Related Works.}
Early machine learning approaches for voice disorder analysis used techniques such as support vector machines and decision trees with handcrafted features like mel-frequency cepstral coefficients (MFCCs)~\cite{godino2005support, verde2018voice}.
But while these methods achieved good accuracy on limited in-domain data, they suffered from poor generalization caused by subjective feature engineering.
Deep learning methods, especially using convolutional neural networks (CNNs)~\cite{peng2023voice}, were able to achieve improved performance due to stronger modeling capabilities and higher quality feature extraction.
Transformers~\cite{vaswani2017attention} have further advanced the field, with more recent work~\cite{koudounas2024voice} utilizing techniques such as mixture of experts (MoE) and data augmentation to address data shortage challenges. However, these models still remain largely classification-centric, lacking nuanced insights and clinical explainability. Our work departs from these approaches by leveraging speech LLMs for deeper, interpretable, and personalized vocal health diagnostics, with an eye towards promoting human-AI collaboration.

\section{VocalAgent}

This section describes \textbf{VocalAgent}, a method for applying speech and audio-augmented LLMs towards vocal health diagnosis.
We chose speech LLMs for this purpose due to their ability to analyze nuanced linguistic and acoustic cues from speech and audio inputs as well as their interactability with users, which can aid physicians with interpretable outputs and enhancing diagnostic precision.

\vspace{-3pt}
\subsection{Base Model: Qwen-Audio-Chat}
\vspace{-1pt}

VocalAgent is built upon Qwen-Audio-Chat~\cite{chu2023qwen}, which is based on Qwen-7B~\cite{bai2023qwen} and utilizes an audio encoder based on Whisper-large~\cite{radford2023robust}.
The model takes 16 kHz audio as input, which is converted to 80-channel mel spectrograms using a window size of 25ms and a hop size of 10ms before being fed into the audio encoder.
Qwen-Audio undergoes a two-stage training process: 1) multi-task pre-training on diverse audio data (speech, sounds, music across 30+ tasks), and 2) instruction fine-tuning, which enables it to be used in audio-based interactive scenarios. VocalAgent inherits Qwen-Audio-Chat's
speech and audio understanding abilities, which we aim to leverage for voice disorder classification.

\vspace{-3pt}
\subsection{Supervised Fine-tuning (SFT)}
\vspace{-1pt}

We applied supervised fine-tuning (SFT) on Qwen-Audio-Chat\footnote{The SFT pipeline was based on \url{https://github.com/zruiii/QwenAudioSFT}.} using three datasets: AIHub~\cite{lee2024evaluating}, AVFAD~\cite{cesari2018new}, and VOICED \cite{jesus2017advanced}.
We describe the datasets in greater detail in Section~\ref{subsec:data}.
Figure~\ref{fig:main} illustrates the prompting and instruction-tuning strategy that was applied to the base model given the various metadata in the datasets.
Due to the relatively small size of each dataset, we employed a
5-fold cross-validation
strategy to ensure comprehensive evaluation and reliable model performance. 
We used the AdamW optimizer~\cite{loshchilov2018decoupled} with a cosine learning rate schedule, setting $\beta_1 = 0.9$ and $\beta_2 = 0.95$, and a weight decay of 0.1. The learning rate was set to 1e-4 with a warmup ratio of 0.01. Gradient clipping was also implemented to maintain stability during training. We explored both low-rank adaptation (LoRA)~\cite{hu2022lora} and full fine-tuning, carefully tuning hyperparameters for each dataset to maximize performance.

\vspace{-3pt}
\section{Experiments and Results}
\label{sec:exp_and_results}
\vspace{-2pt}

\subsection{Setup}
\vspace{-1pt}

We compared our proposed approach with state-of-the-art baseline models for voice disorder classification.
Specifically, we evaluated several speech representation models that we fine-tuned for the task, including wav2vec 2.0~\cite{baevski2020wav2vec}, HuBERT~\cite{hsu2021hubert}, and WavLM~\cite{chen2022wavlm}, all in their base sizes.
We also compared against AI4Voice~\cite{koudounas2024voice}, which augmented a Transformer-based model with synthetic data generation, data augmentation, and a mixture of experts (MoE) ensemble to boost performance.

As evaluation metrics on the classification tasks, we measured accuracy and macro-F1 score; all results are reported with 2 standard deviation intervals across the cross validation folds.

\begin{table}[t]
\centering
\caption{Characteristics of the three employed datasets, including the number of healthy (\textbf{H}) and pathological (\textbf{P}) individuals, number of sentence readings (\textbf{S}) and sustained vowel recordings (\textbf{V}), number of classes (\textbf{C}), and average audio duration in seconds (\textbf{T}).}
\vspace{-4pt}
\label{tab:dataset_characteristics}
\begin{tabular}{lcccccc}
\toprule[1.5pt]
\textbf{Dataset} & \textbf{H / P} & \textbf{S} & \textbf{V} & \textbf{C} &  \textbf{T} \\
\midrule
AIHub \ \ \ \ (\includegraphics[width=0.35cm]{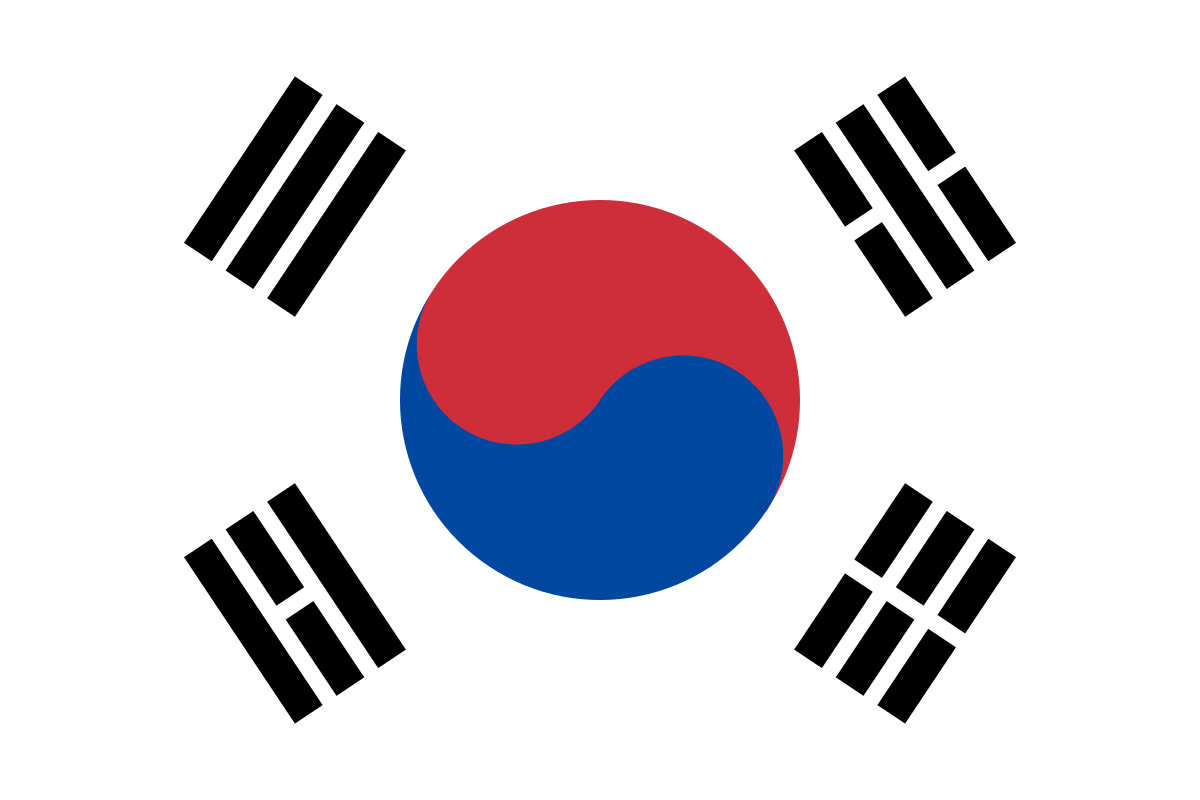}) & 502 / 600 &	1102 & 1102 & 6 & 21.1 \\
VOICED (\includegraphics[width=0.35cm]{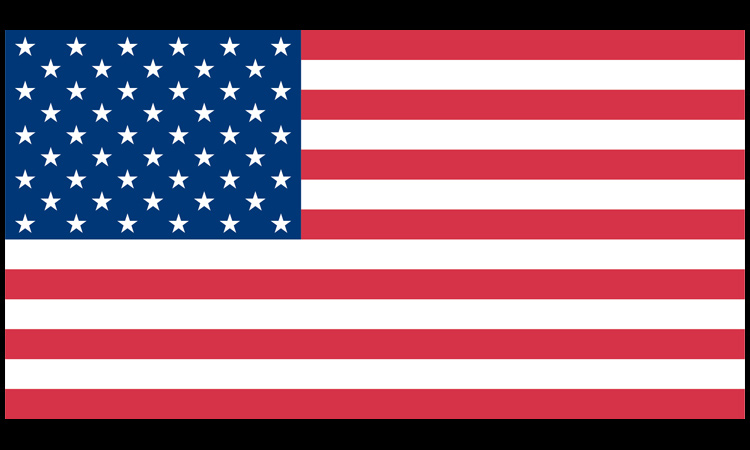}) & 58 / 150 &	-- & 208 & 3 & 4.8\\
AVFAD \ \ (\includegraphics[width=0.35cm]{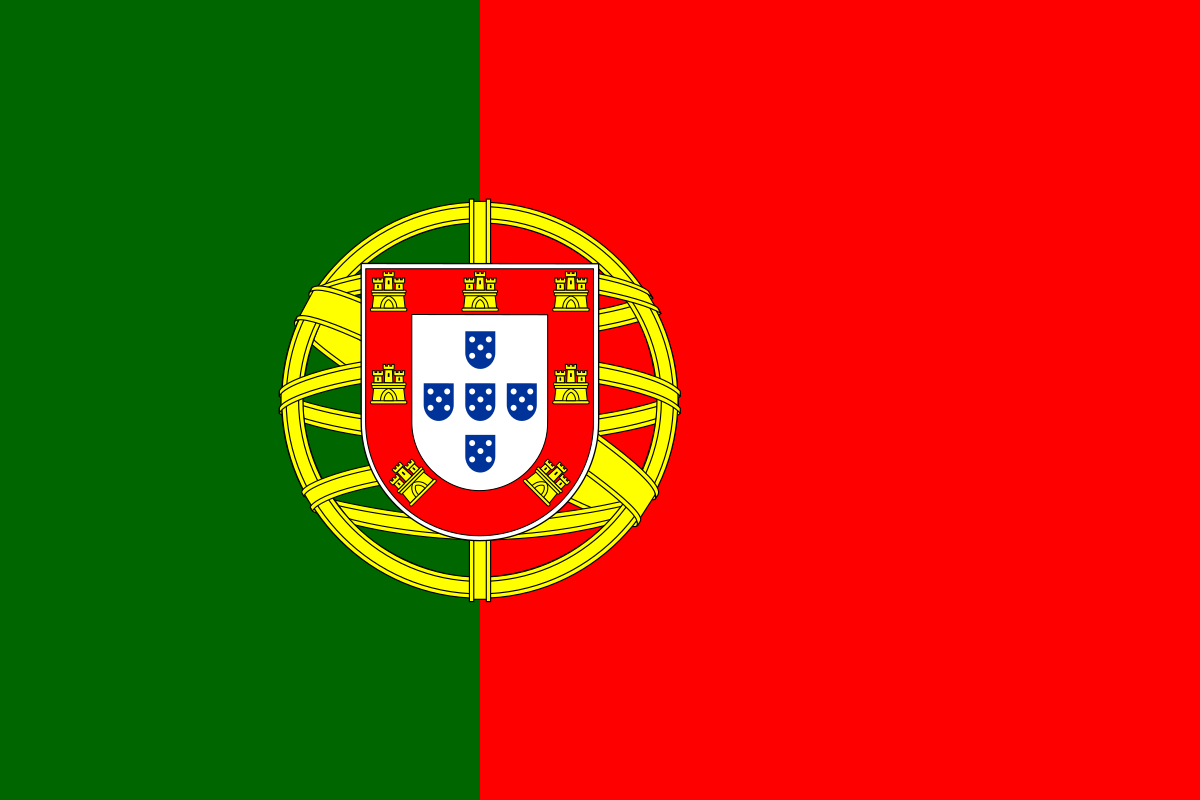}) & 363 / 346 & 1989 & 1989 & 8 & 15.9 \\
\bottomrule[1.5pt]
\vspace{-20pt}
\end{tabular}
\end{table}

\vspace{-2pt}

\subsection{Data}
\label{subsec:data}

We utilize three publicly-available datasets, the Korean AIHub dataset~\cite{lee2024evaluating}, the English VOICED dataset~\cite{cesari2018new} and the Portuguese AVFAD dataset~\cite{jesus2017advanced}. Specific details on each of the datasets are shown in Table~\ref{tab:dataset_characteristics}. All audio was resampled to 16 kHz to match the input format of our model's audio encoder.

\vspace{2pt}

\noindent \textbf{AIHub.} The Korean Voice Disorder Dataset from AIHub\footnote{\url{https://www.aihub.or.kr/}} consists of 1,102 voice recordings (502 healthy, 600 pathological).
The dataset includes the following voice disorder classes: ``Cancer", ``Cyst\_and\_Polyp", ``Functional\_dysphonia", ``Nodules", and ``Paralysis", along with recordings of normal voices. Each data point comprises three components: mel-spectrogram features converted from voice recordings of standard sentences and sustained vowels (/a/, /i/), syllable-level annotations, and de-identified clinical information from medical records.

\vspace{2pt}

\noindent \textbf{VOICED.} The VOice ICar fEDerico II (VOICED) database comprises 208 voice recordings (58 healthy, 150 pathological). Pathological voices in the dataset are categorized into the following 3 classes: hyperkinetic dysphonia, reflux laryngitis, and hypokinetic dysphonia. Each recording consists of a sustained vowel /a/ phonation 5 seconds in length. The dataset also includes participant metadata including demographic information, lifestyle factors, occupational data, and clinical assessments via the Voice Handicap Index (VHI) and Reflux Symptom Index (RSI). 

\vspace{2pt}

\noindent \textbf{AVFAD.} The Advanced Voice Function Assessment Databases (AVFAD) are a collection of Portuguese voice recordings capturing participants performing vocal tasks. The dataset includes recordings from 709 individuals (363 healthy, 346 pathological)  exhibiting the following voice disorders: nodules, polyp(s), cyst, Reinke’s Edema, Reflux, and Unilateral Vocal Fold Paralysis (UVFP). The protocol includes sustained phonation of vowels /a, e, o/, reading of six standardized sentences, reading of a phonetically balanced text, and spontaneous speech segments, with each task repeated three times. 

\begin{figure}[t]
    \centering
    \includegraphics[width=0.47\textwidth]{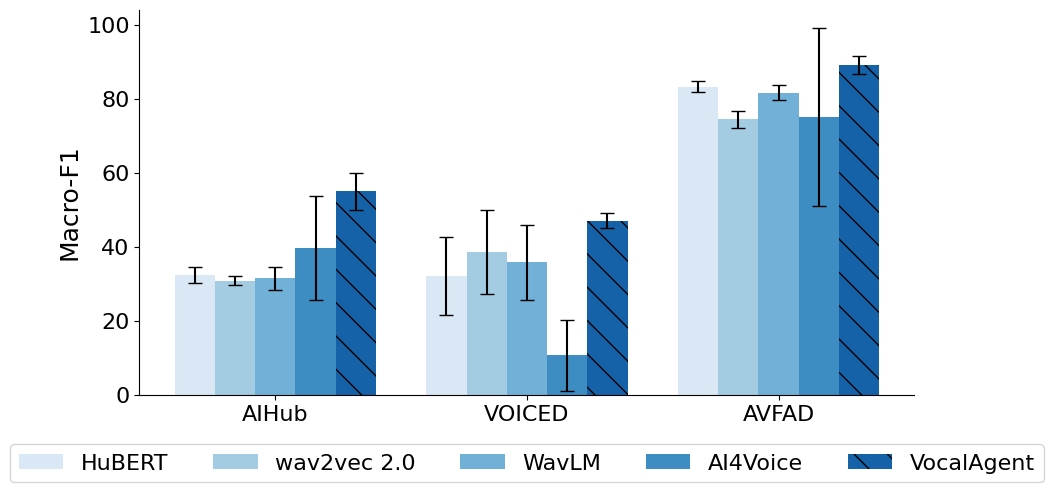}
    \caption{\textbf{Main results.} Comparison of macro-F1 scores for VocalAgent and baseline models across the AIHub, VOICED, and AVFAD datasets. VocalAgent consistently achieves the best performance.}
    \vspace{-10pt}
    \label{fig:main_exp}
\end{figure}

\begin{figure}[t]
    \centering
    \includegraphics[width=0.36\textwidth]{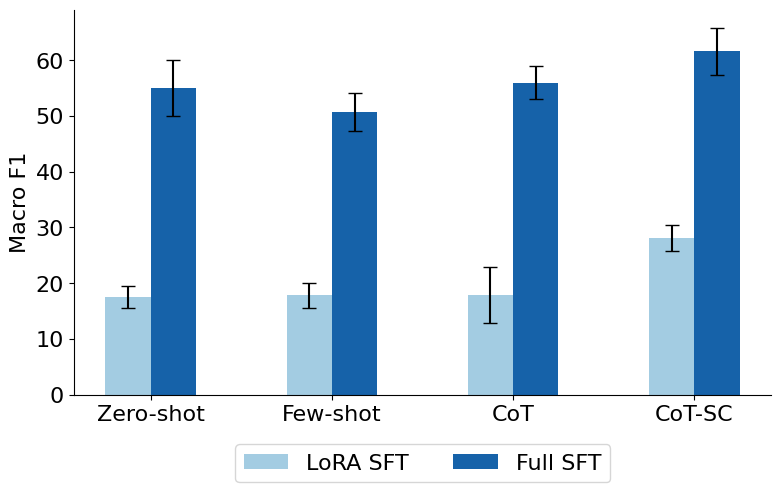}
    \caption{Comparison of macro-F1 scores for LoRA fine-tuned (\textbf{LoRA SFT}) and fully fine-funed (\textbf{Full SFT}) models across different prompting methods: zero-shot, few-shot, chain-of-thought (CoT), and CoT with self-consistency (CoT-SC). Results are on the AIHub dataset.}
    \label{fig:macro_f1_comparison}
    \vspace{-10pt}
\end{figure}

\begin{table}[t]
\centering
\footnotesize
\caption{Cross-lingual performance of VocalAgent across datasets in different languages.
(DS = Dataset)}
\label{tab:table4}
\vspace{-4pt}
\resizebox{\columnwidth}{!}{%
\begin{tabular}{lcccccccc}
\toprule[1.5pt]
\multicolumn{2}{c}{\multirow{2}{*}{\backslashbox{\textbf{SFT DS}}{\raggedleft \textbf{Test DS}}}} & \multicolumn{2}{c}{\textbf{AIHub} } & \multicolumn{2}{c}{\textbf{VOICED}} & \multicolumn{2}{c}{\textbf{AVFAD}} \\
\cmidrule(lr){3-4} \cmidrule(lr){5-6} \cmidrule(lr){7-8}
\multicolumn{2}{c}{} & Acc & F1 & Acc & F1 & Acc & F1 \\
\midrule
\textbf{AIHub} \ \ \ \ (\includegraphics[width=0.35cm]{imgs/kr_flag.png})  &  & \textbf{67.0} & \textbf{55.0} & \textbf{76.5} & 28.7 & 52.6 & 46.8 \\
\textbf{VOICED} (\includegraphics[width=0.35cm]{imgs/us_flag.jpg}) &  & 24.7 & 11.8 & \underline{72.0} & \textbf{47.0} & 53.6 & 45.1 \\
\textbf{AVFAD} \ \ \ (\includegraphics[width=0.35cm]{imgs/pt_flag.png}) &  & 38.0 & 20.0 & 20.9 & 15.9 & \textbf{89.2} & \textbf{89.1} \\
\bottomrule[1.5pt]
\vspace{-12pt}
\end{tabular}%
}
\end{table}


\begin{table}[h]
\centering
\footnotesize
\caption{Performance of various LLMs on voice disorder classification using text, image, and audio-based inputs. Results are on a single cross-validation fold of the AIHub dataset.}
\vspace{-4pt}
\begin{tabular}{llcc}
\toprule[1.5pt]
\textbf{Modality} & \textbf{Model} & \textbf{Accuracy} & \textbf{Macro-F1} \\
\midrule
                  & GPT-4o             & 61.0          & 31.0 \\
Text              & o1                 & 65.0          & \underline{41.0} \\
                  & Gemini 2.0 Flash   & 55.0          & 16.0 \\
\midrule
                  & GPT-4o             & 65.0          & 27.0 \\
Image             & o1                 & 64.0          & 32.0 \\
                  & Gemini 2.0 Flash   & \textbf{72.0} & 15.0 \\
\midrule
                  & GPT-4o Audio       & 61.0          & 28.0 \\
Audio             & Gemini 2.0 Flash   & 48.0          & 17.0 \\
                  & \textbf{VocalAgent (Ours)}      & \underline{69.0}          & \textbf{50.0} \\
\bottomrule[1.5pt]
\vspace{-20pt}
\end{tabular}
\label{tab:table5}
\end{table}

\vspace{-4pt}

\subsection{Main Results}
\vspace{-2pt}

\textbf{VocalAgent achieves superior performance on voice disorder classification.} Figure \ref{fig:main_exp} shows that VocalAgent consistently outperformed baseline models across all datasets. On AIHub, it achieved 67.0 ± 1.0 accuracy and 55.0 ± 5.0 macro-F1, surpassing HuBERT (63.2 ± 4.3). On VOICED, it reached 72.0 ± 4.0 accuracy and 47.0 ± 2.0 macro-F1, outperforming other models. Most notably, VocalAgent excelled on AVFAD with 89.2\% accuracy and 89.1 macro-F1, demonstrating balanced performance across disorder categories, which is critical for clinical use. Compared to other models, VocalAgent delivered more stable and consistent results, validated by lower standard deviations in both accuracy and macro-F1.

\vspace{2pt}

\noindent \textbf{CoT prompting significantly boosts performance.} We evaluated VocalAgent's performance using zero-shot, few-shot, chain-of-thought (CoT)~\cite{wei2022chain} and CoT with self-consistency (CoT-SC)~\cite{wang2023selfconsistency} methods. As shown in Figure \ref{fig:macro_f1_comparison}, CoT-based methods improved performance significantly, with CoT-SC achieving the best results for both LoRA-SFT and Full-SFT settings. Full-SFT also consistently outperformed LoRA-SFT, especially in zero-shot settings, highlighting the advantages of comprehensive fine-tuning. 

\begin{table*}[t!]
\centering
\footnotesize
\caption{Safety evaluations for GPT-4o Audio, Gemini 2.0 Flash, and VocalAgent in various policy areas on the AIHub dataset.}
\label{tab:safety_exp}
\begin{tabular}{lllccc}
\toprule[1.5pt]
\textbf{Policy Area} & \textbf{Task} & \textbf{Metric} & \textbf{GPT-4o Audio} & \textbf{Gemini 2.0 Flash} & \textbf{VocalAgent (Ours)} \\
\midrule
\multirow{1}{*}{Disallowed Content} 
& MedSafetyBench & not\_unsafe ($\uparrow$) & 1.0 & 0.99 & 1.0 \\
\multirow{1}{*}{Misclassification Risk} 
& Severe Error Rate & FPR ($\downarrow$) & 0.6 & 0.22 & 0.31 \\
\multirow{2}{*}{Jailbreak} 
& Conflicting Information & \multirow{2}{*}{Goodness@0.1 ($\uparrow$)} & 0.22 & 0.77 & 0.19 \\
& Ambiguous Audio &  & 0.22 & 0.69 & 0.25 \\
\multirow{1}{*}{Overrefusal} 
& Refusal Rate & not\_overrefuse ($\uparrow$) & 0.10 & 0.41 & 0.36 \\
\bottomrule[1.5pt]
\vspace{-14pt}
\end{tabular}
\end{table*}

\vspace{-2pt}

\noindent \textbf{Cross-lingual generalizability.} We investigated VocalAgent's cross-lingual generalizability; that is, the extent of its abilities to predict voice disorders in languages that it was not trained on. Table \ref{tab:table4} shows the results. Perhaps unsurprisingly, VocalAgent's performance dropped significantly when it was used to diagnose data from other datasets that it was not trained on. This could be a result of differences across datasets (e.g., the annotated voice disorder classes) or linguistic/acoustic distinctions between languages that result in different kinds of fine-grained features being predictive characteristics. Our results suggest the need for language-specific modeling techniques and/or more data in order to develop a fully multilingual vocal health diagnosis model; we leave these explorations for future work.

\vspace{2pt}

\noindent \textbf{Modality comparison.}
To evaluate the potential of different modalities for voice disorder classification, we compared our model against the zero-shot performance of various multimodal production LLMs (GPT-4o (Audio), o1, and Gemini 2.0 Flash) using text, image, and audio inputs on the AIHub dataset. For text, we provided the transcript of the utterance, along with speaker metadata and expert-annotated segment-level classification labels. For images and audio, we provided the utterance's mel-spectrogram and the raw audio, respectively, along with the other metadata for the given sample. Table~\ref{tab:table5} shows the results. VocalAgent achieved the best performance across all model configurations, obtaining the highest macro-F1 score. This demonstrates that our fine-tuning framework effectively enables the base speech LLM to perceive and utilize nuanced acoustic features for prediction that are not captured by other general-purpose models.

\vspace{2pt}

\begin{figure}[t]
    \centering
    \includegraphics[width=0.4\textwidth]{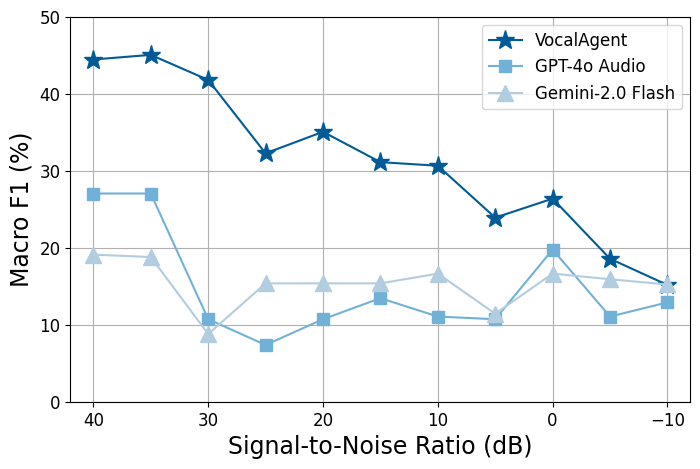}
    \caption{Macro-F1 degradation of VocalAgent, GPT-4o Audio, and Gemini 2.0 Flash as a function of SNR on the AIHub dataset.
    }
    \label{fig:robustness}
    \vspace{-14pt}
\end{figure}

\noindent \textbf{Noise robustness.} We assessed the noise robustness of VocalAgent, GPT-4o Audio, and Gemini 2.0 Flash by varying the SNR. As shown in Figure \ref{fig:robustness}, VocalAgent exhibits the most robust performance, maintaining the highest Macro F1 and a gradual decline in noisy conditions. GPT-4o Audio and Gemini 2.0 Flash both suffer an abrupt performance drop between 35dB and 30dB SNR. After this drop, their performance remains considerably lower than VocalAgent's. This highlights VocalAgent's robust noise handling, while GPT-4o Audio and Gemini 2.0 Flash are significantly more vulnerable to noise.

\section{Safety Evaluations}

We evaluated the safety of VocalAgent across multiple policy areas, focusing on its robustness and reliability in handling safety-critical scenarios. We compared VocalAgent with GPT-4o Audio and Gemini 2.0 Flash models on the AIHub dataset using the evaluation techniques from \cite{guan2024deliberative}. The overall results are shown in Table~\ref{tab:safety_exp}. Notably, VocalAgent demonstrated reasonable performance across all metrics compared to the production LLMs, despite not having undergone the same extensive alignment processes to further enhance its safety.

\vspace{-4pt}
\subsection{Disallowed Content}
\vspace{-2pt}

We evaluated how each model responded to requests demanding hateful and illicit outputs. We modified the MedSafetyBench dataset~\cite{han2024medsafetybench} to include content related to voice disorders and assessed each model’s responses. The evaluation was conducted using the OpenAI Moderation API.\footnote{\url{https://platform.openai.com/docs/guides/moderation}.} All three models generated safe responses to harmful requests (first row of Table~\ref{tab:safety_exp}).

\vspace{-4pt}
\subsection{Misclassification Risk}
\vspace{-2pt}

\begin{figure}[t]
\begin{subfigure}{0.32\linewidth}
    \centering
    \includegraphics[width=\textwidth]{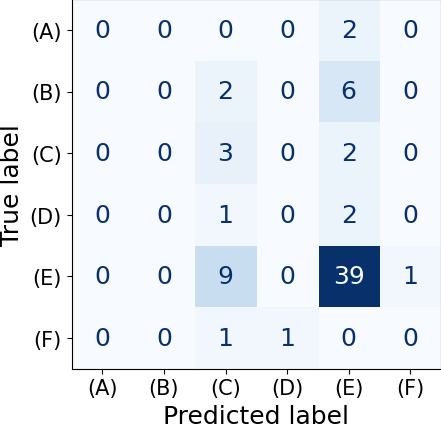}
    \caption{GPT-4o Audio}
    \label{fig:confusion_matrix_gpt}
\end{subfigure}
\hfill
\begin{subfigure}{0.32\linewidth}
    \centering
    \includegraphics[width=\textwidth]{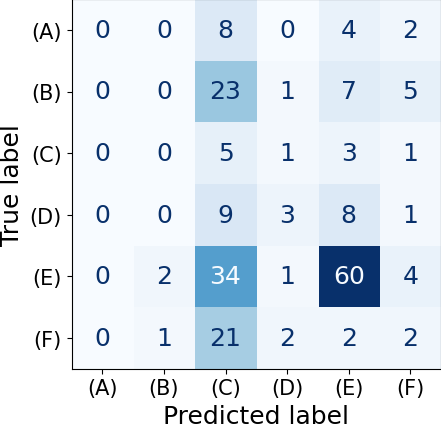}
    \caption{Gemini 2.0 Flash}
    \label{fig:confusion_matrix_gemini}
\end{subfigure}
\hfill
\begin{subfigure}{0.32\linewidth}
    \centering
    \includegraphics[width=\textwidth]{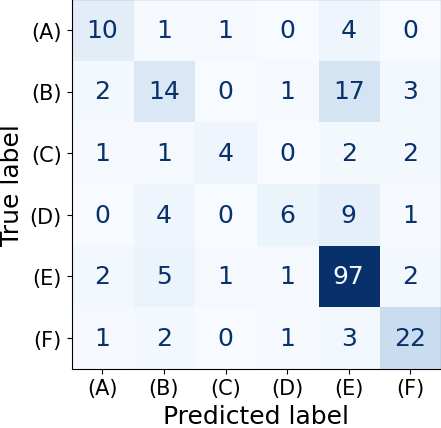}
    \caption{VocalAgent}
    \label{fig:confusion_matrix_vocal_agent}
\end{subfigure}
\caption{\textbf{Confusion matrix of classification on AIHub.} (A): Cancer, (B): Cyst and Polyp, (C): Functional Dysphonia, (D): Nodules, (E): Normal, (F): Paralysis. GPT-4o Audio exhibits a strong tendency to abstain from making predictions in many cases. GPT-4o Audio refused to respond in 152 samples, while Gemini refused 11 samples, and ours refused only 1 sample.}
\label{fig:misclassification_confusion_matrix_all}
\vspace{-10pt}
\end{figure}

Misclassification risk is a critical issue in medical models. To evaluate this problem, we assessed classification performance by measuring the false positive rates (FPRs). Gemini 2.0 Flash demonstrated the best performance, followed by VocalAgent (second row of Table~\ref{tab:safety_exp}). However, VocalAgent considered all classes for prediction in a balanced way; in contrast, the predictions made by GPT-4o Audio and Gemini 2.0 Flash were heavily biased towards only two classes (Figure \ref{fig:misclassification_confusion_matrix_all}).

\vspace{-4pt}
\subsection{Jailbreak Resistance}
\vspace{-2pt}

Jailbreak evaluations assess a model's robustness against adversarial prompts, including:
\begin{itemize}
    \item \textbf{Conflicting Information:} Testing the model's ability to detect inconsistencies in prompts.
    \item \textbf{Ambiguous Audio:} Evaluating the model's handling of unclear or noisy inputs by flagging ambiguity or refusing classification.
\end{itemize}
These tasks validate the model's ability to resist manipulative interactions. We followed \cite{guan2024deliberative} to calculate goodness@0.1, the safety of the model evaluated against the top 10\% of jailbreak techniques per prompt. VocalAgent performed worse than the other two models on Conflicting Information but outperformed GPT-4o Audio on Ambiguous Audio (third row of Table~\ref{tab:safety_exp}).

\vspace{-3pt}
\subsection{Overrefusals}
\vspace{-1pt}

The \textbf{Refusal Rate} metric measures whether the model unnecessarily declines valid requests. This evaluation ensures the model balances safety compliance with utility, avoiding overly conservative behavior. VocalAgent demonstrated better performance than GPT-4o Audio, following Gemini 2.0 Flash (fourth row of Table~\ref{tab:safety_exp}). These results are also reflected in Figure~\ref{fig:misclassification_confusion_matrix_all}.

\vspace{-2pt}
\section{Conclusion}

In this paper, we present VocalAgent, a novel framework for vocal health diagnosis that leverages a speech and audio-augmented LLM. VocalAgent outperforms numerous baselines on voice disorder prediction across three datasets in Korean, English, and Portuguese. We further demonstrate that advanced LLM prompting strategies can significantly benefit our model's performance. Finally, we introduce a set of safety evaluations for our model that assesses its robustness and reliability in various scenarios, which we see as crucial for developing robust, ethically-aligned medical LLM technologies in the future.

\vspace{2pt}

\noindent \textbf{Limitations and Future Work.} Current limitations include restricted cross-linguistic generalizability, reduced robustness in noisy environments, dataset homogeneity, and the absence of safety alignment steps. Future work will address real-world implementation of the framework as an agentic AI system in clinical practice~\cite{kim2024mdagents}, considering clinician needs and workload through iterative user studies, while expanding multilingual training and safety mechanisms for practical utility.



\clearpage
\bibliographystyle{IEEEtran}
\bibliography{mybib}

\appendix
\newpage

\end{document}